%% file: ms.tex
\documentclass[preprint2]{aastex}
\shorttitle{Polarimetry of HH 1--2 Region}
\shortauthors{Kwon et al.}

\usepackage{times}
\usepackage{threeparttable} 

\slugcomment{To appear in the Astrophysical Journal Letters}

\begin{document}

\fontsize{10}{10.6}\selectfont

\title{NEAR-INFRARED CIRCULAR POLARIZATION SURVEY IN STAR-FORMING REGIONS: CORRELATIONS AND TRENDS}

\author{\sc Jungmi Kwon\altaffilmark{1,2,7},
            Motohide Tamura\altaffilmark{1,2}, James H. Hough\altaffilmark{3},
            Nobuhiko Kusakabe\altaffilmark{2}, Tetsuya Nagata\altaffilmark{4},
            Yasushi Nakajima\altaffilmark{5}, Phil W. Lucas\altaffilmark{3}, 
            Takahiro Nagayama\altaffilmark{6}, and Ryo Kandori\altaffilmark{2}}
\affil{$^1$ Department of Astronomy, Graduate School of Science, The University of Tokyo, 7-3-1 Hongo, Bunkyo-ku, Tokyo 113-0033, Japan; jungmi.kwon@astron.s.u-tokyo.ac.jp}
\affil{$^2$ National Astronomical Observatory of Japan, 2-21-1 Osawa, Mitaka, Tokyo 181-8588, Japan}
\affil{$^3$ Centre for Astrophysics Research, University of Hertfordshire, College Lane, Hatfield AL10 9AB, UK}
\affil{$^4$ Department of Astronomy, Kyoto University, Kyoto 606-8502, Japan}
\affil{$^5$ Center of Information and Communication Technology, Hitotsubashi University, 2-1 Naka, Kunitachi, Tokyo 186-8601, Japan}
\affil{$^6$ Department of Astrophysics, Kagoshima University, 1-21-35 Korimoto, Kagoshima 890-0065, Japan}
\affil{$^7$ JSPS Postdoctoral Fellow}

\begin{abstract}

\fontsize{10}{10.6}\selectfont

We have conducted a systematic near-infrared circular polarization (CP) survey in star-forming regions, covering high-mass, intermediate-mass, and low-mass young stellar objects. All the observations were made using the SIRPOL imaging polarimeter on the Infrared Survey Facility 1.4 m telescope at the South African Astronomical Observatory.  
We present the polarization properties of ten sub-regions in six star-forming regions. 
The polarization patterns, extents, and maximum degrees of linear and circular polarizations 
are used to determine the prevalence and origin of CP in the star-forming regions. 
Our results show that the CP pattern is quadrupolar in general, 
the CP regions are extensive, up to 0.65 pc, 
the CP degrees are high, up to 20 \%, 
and the CP degrees decrease systematically from high- to low-mass young stellar objects.  
The results are consistent with dichroic extinction mechanisms 
generating the high degrees of CP in star forming regions.
\end{abstract}


\section{INTRODUCTION}

    Polarimetry is an important tool for studying the physical processes in the interstellar medium, 
including star-forming regions (see, e.g., Whittet 1992). 
The polarimetry of young stellar objects (YSOs) and their circumstellar structures 
provides invaluable information about the distribution of matter and the configuration of magnetic fields
in their environments. 
The polarization of YSOs is mainly caused by the scattering and absorption of light by dust particles. 
After scattering and absorption, the light is elliptically polarized 
and thus its polarization state must be determined by measuring both the linear polarization 
(LP; Stokes $Q$ and $U$) and the circular polarization (CP; Stokes $V$) components.  

Early attempts to measure CP were made at optical wavelengths 
to obtain information about grain material in the interstellar medium (e.g., Martin 1972) 
or about multiple scattering processes (M\'{e}nard et al. 1988).  
However, because of heavy extinction by dust in star-forming regions, 
optical measurements of CP have been very limited (Clayton et al. 2005) 
in contrast to the many observations of LP in both the optical and infrared (see e.g., Bastien 2011).
It is therefore important to measure CP at infrared wavelengths for a large number of star-forming regions.
Since many of the associated reflection nebulae are known to be very extensive (more than a few arcminutes)
and highly linearly polarized (more than 10 \%),
CP observations have to be made with care to avoid any of the LP being measured as CP.

In this Letter, we present the results of a first systematic CP survey in star-forming regions. 
In Section 2, we describe the observations and data reduction. 
In Section 3, we present the results of the CP survey in ten sub-regions of six star-forming regions. 
In Section 4, we discuss the origin of the CP. 
This is one of the most extensive polarimetric studies in star-forming regions, 
showing how aligned grains are important in massive star-forming regions.

\section{OBSERVATIONS AND DATA REDUCTION}

All observations were carried out using the SIRPOL imaging polarimeter 
on the Infrared Survey Facility (IRSF) 1.4 m telescope at the South African Astronomical Observatory.  
SIRPOL consists of a single-beam polarimeter and an imaging camera (Nagayama et al. 2003), SIRIUS, 
which has three 1024 $\times$ 1024 HgCdTe IR detector arrays.  
SIRPOL can be used for wide-field (7\farcm7 $\times$ 7\farcm7 with a scale of 0\farcs45 pixel$^{-1}$) imaging polarimetry 
in the $J$, $H$, and $Ks$ bands, simultaneously (Kandori et al. 2006). 
The polarimetry unit consists of a stepped achromatic half-wave (for linear polarimetry) 
or an achromatic quarter-wave (for circular polarimetry) plate followed by a high-efficiency polarizer. 
In the CP mode of SIRPOL a continuous rotating half-wave plate, upstream of the quarter-wave plate, suppresses any linear-to-circular polarization conversions by effectively smearing out the LP (also see Chrysostomou et al. 1997).  

Six star-forming regions were surveyed in both the CP and LP modes from 2005 to 2007: OMC-1, NGC 6334-V, OMC-2, OMC-3, Cha IRN, and L1551.  
In the CP modes, we performed 10-s exposures at two wave-plate angles, 
0$\arcdeg$ and 90\arcdeg, at ten dithered positions for each set.  
In order to obtain a better signal-to-noise ratio, the observation sets were repeated 
toward the target object and the sky backgrounds, 
nine times for OMC-1, NGC 6334-V, and L1551, three times for Cha IRN, and six times for the others. 
In the LP mode, we performed 10-s exposures at four wave-plate angles, 
0\arcdeg, 45\arcdeg, 22\fdg5, and 67\fdg5, at 10 dithered positions for each set.  
The observation sets were repeated ten times toward the target object and the sky backgrounds, except for Cha IRN, which was repeated three times. 
The data were processed using IRAF in the standard manner, 
which included dark-field subtraction, flat-field correction, sky-bias subtraction, and frame registration. 
Further processing of the data, such as removing artificial stripe patterns and performing registrations, 
was performed as described by Kwon et al. (2011).  
Polarimetry of extended sources was carried out on the combined intensity images for each exposure cycle 
(a set of exposures at four wave plate angles at the same dithered position).  
The Stokes parameters $I$, $Q$, $U$, and $V$ were calculated as described by Kwon et al. (2013). 
Then, the obtained images of CP degrees were generated by 3 $\times$ 3 pixel smoothing.

In this Letter, we designate individual infrared reflection nebula(e) as IRN(e).  
We also designate the IRN associated with an IR source X as IRN (X), such as IRN (IRc2), 
rather than numbering each. Exceptions to this naming scheme are made for NGC 6334-IV IRN and NGC 6334-V IRN. 
NGC 6334-IV IRN is a small nebula detected in the NGC 6334-V field but that actually belongs to NGC 6334-IV, 
which is located to the northeast of NGC 6334-V.  
The illuminating source of NGC 6334-IV IRN is invisible. 
In addition, there is considerable uncertainty over the illuminating source of NGC 6334-V IRN. 
Because the illuminating sources of these two IRNe are obscure, we refer to them by the more specific names of NGC 6334-V IRN and NGC 6334-IV IRN rather than by referencing their sources.

\section{RESULTS}
\subsection{Polarization Properties in Star-forming Regions}

In this section, we briefly describe the CP properties of the IRNe around various YSOs, 
ranging from high-mass to low-mass YSOs.  
Our sample was selected from one or more IRNe believed to be associated with outflows from embedded YSOs.
Figures 1 and 2 show Stokes $I$ and $V$ images of each survey region from the IRSF/SIRPOL observations, respectively.
These are the first CP imaging data for OMC-2, OMC-3, and L1551. 
Note that since a number of the lower mass star-forming regions show only very faint CP patterns, 
only upper limits are obtained for the nebulae.

\paragraph{OMC-1} 
We have confirmed a region of CP in the OMC-1 field, showing a  quadrupolar pattern, 
as found in previous studies (Bailey et al. 1998; Buscherm\"{o}hle et al. 2005; Fukue et al. 2009).  
The extent of the CP is $\sim 210\arcsec$ (0.45 pc).
The deep CP image indicates that the quadrupolar pattern is centered on IRc2, 
suggesting that this young star or cluster, the most massive in the region, 
is the dominant source responsible for the large CP. 
Note that one of the negative patterns near the Trapezium is less distinct than the other patterns. 
No clear pattern is observed around the Becklin--Neugebauer object, 
although this source is embedded in the negative CP region. 
The CP degree ranges from $+16.2 \pm 0.2$ \% to $-4.3 \pm 0.2$ \% 
in the $K_s$ band.

\paragraph{NGC 6334-V} 
We have confirmed two regions of CP in the NGC 6334-V field.  
One is seen in the NGC 6334-V IRN, while the other is seen around NGC 6334-IV (see the next paragraph).
The NGC 6334-V IRN shows a region of CP extending over $\sim 80\arcsec$ (0.65 pc, Kwon et al. 2013), 
which includes an additional faint region with high negative CP ($-11$ \%) 
located $\sim 40\arcsec$ to the west.
It is larger than the CP region in OMC-1 (210$\arcsec$ equivalent to 0.45 pc at 450 pc). 
The CP degree ranges from $+19.8 \pm 0.5$ \% to $-9.9 \pm 0.1$ \% 
in the $K_s$ band.

\paragraph{NGC 6334-IV} 
NGC 6334-IV IRN does not show a clear quadrupolar structure 
of positive and negative degrees of CP as found for NGC 6334-V IRN, 
however, it does show a faint negative bipolar pattern. 
The CP degree detected in the northern nebular lobe of  NGC 6334-IV IRN is 
$-3.5 \pm 0.1$ \%
in the $K_s$ band.

\paragraph{OMC-2} 
We have discovered a somewhat complex CP region in the OMC-2 field 
including illuminating sources IRS 1--IRS 4. 
In the OMC-2 region, the largest nebula, OMC-2 IRN (IRS 1), 
shows a faint quadrupolar pattern in its eastern nebular lobe. 
The positive and negative pattern is seen in its eastern nebular lobe like half a butterfly. 
Note that the LP image of this IRN itself is similarly monopolar. 
The detected CP in the $K_s$ band lies mostly in the range 
$+4.6 \pm 0.4$ \% 
to $-2.1 \pm 0.3$ \%. 
Both OMC-2 IRN (IRS 2) and OMC-2 IRN (IRS 4) show faint positive monopolar patterns. 
The positive CP degree detected in the western nebular lobe of OMC-2 IRN (IRS 2) is $+1.8 \pm 0.2$ \% 
in very small areas ($\sim 20\arcsec$). 
The positive CP degree detected in the northern nebular lobe of OMC-2 IRN (IRS 4) is 
$+4.3 \pm 0.2$ \% 
with an extent of $\sim 40\arcsec$. 
Since OMC-2 IRN (IRS 2) and OMC-2 IRN (IRS 4) are very close to each other, 
some distortion of the more usual LP or CP patterns is expected 
for overlapping nebular regions.

\paragraph{OMC-3} 
We have discovered a region of CP around IRAS 05329-0505 in the OMC-3 field. 
OMC-3 IRN (IRAS 05329-0505) shows a faint positive bipolar pattern with an extent of 100$\arcsec$ 
(equivalent to 0.22 pc at 450 pc). 
It also shows larger degrees of CP near the illuminating source, 
decreasing with distance from the source, 
and with a clearer detection in the eastern nebular lobe than in the western lobe. 
The degree of detected CP is $+1.9 \pm 0.2$ \% 
in the $K_s$ band.

\paragraph{Cha IRN} 
We have confirmed a region of CP in the Cha IRN field 
showing an asymmetric quadrupolar pattern (Gledhill et al. 1996).  
The sign reversal in the Stokes $V$ between quadrants centered on the source position can be clearly seen,
corresponding to changes in the handedness of CP. 
The CP degrees detected in the eastern nebular lobe 
lie mostly in the range $+1.1 \pm 0.1$ \% 
to $-0.8 \pm 0.1$ \% in the $H$ band. 
This result is consistent with the previous CP of the Cha IRN (Gledhill et al. 1996) 
but with a higher sensitivity.
Note that since the CP of the Cha IRN shows clearer pattern in the $H$ band than in the $K_s$ band, exceptionally, we describe the CP of the Cha IRN in the $H$ band.

\paragraph{L1551} 
We have discovered two CP regions around IRS 5 and IRS 5 NE in the L1551 field, 
although with low signal to noise. 
The CP of L1551 IRN (IRS 5) shows a monopolar pattern as expected from their morphologies 
based on their intensities and linear polarimetry, 
i.e., the counterpart is invisible due to heavy extinction by the circumstellar disk. 
The CP extent is 30$\arcsec$ (equivalent to 0.02 pc at 150 pc).  
The CP degree detected in the south-western nebular lobe of L1551 IRN (IRS 5) is 
$+1.6 \pm 0.2$ \%
in the $K_s$ band.  
In the case of L1551 IRN (IRS 5 NE), the CP pattern is not clear. 
The CP degree in the western nebular lobe of L1551 IRN (IRS 5 NE) is $-1.1 \pm 1.0$ \% 
in the $K_s$ band, 
and we suggest, therefore, that the CP has an upper limit of $\sim 2$ \% in L1551 IRN (IRS 5 NE).

\section{DISCUSSION}

Based on the description in the previous section, we summarize each star forming region in Table 1, 
showing the illuminating sources, the polarization patterns, their extents, and the maximum degrees 
of both CP and LP.  The CP results are as follows: 
(1) CP is commonly detected in star-forming regions; 
(2) the extents of the infrared CP region are quite large ($\sim 0.1$--$1$ pc); 
(3) the maximum degrees of CP are high, up to 20 \%, and generally seem to be highest in the $K_s$ band. 
In addition, we summarize the LP results; 
(4) IRNe associated with the various YSOs have either bipolar or monopolar morphology, 
and occasionally they are asymmetric around each illuminating source; 
(5) the extents of the regions of LP are also large, ranging from $\sim 0.1$ pc to $\sim 1$ pc, 
with an average of $\sim 0.3$ pc;  
(6) the maximum degrees of LP are very large, ranging from $\sim 20$ \% to $\sim 70$ \% 
(an average of $\sim 40$ \%), close to the theoretical maximum of 90 \% (Kwon et al. 2013).
Our survey shows that large ($> 2$ \%) infrared CPs are universally observed in the reflection nebulae in star forming regions, which is in contrast to the results (CP $< 1.5$ \%) of the previous aperture infrared polarization survey (Lonsdale et al. 1980).

Large degrees of CP in star forming regions are produced either by dichroic scattering 
(i.e., scattering off aligned grains) or by dichroic extinction of initially linearly polarized light, 
with the latter mechanism thought to be the most likely (Lucas et al. 2005, see also Kwon et al. 2013), 
as suggested for OMC-1, HH 135, and NGC 6334.  
Multiple scattering by spherical or by non-aligned dust grains 
cannot produce the high degrees of CP observed. 
Although detailed modeling of each star-forming region is beyond the scope of this Letter, 
we have confirmed that the $shadow.f$ scattering code of Lucas et al. (2004), 
assuming dichroic scattering (see Kwon et al. 2013), can reproduce the observed CP 
ranging from $< 1$ \%  to $\sim 20$ \%, with Av = $0$--$\sim 40$ mag, 
as well as the quadrupolar patterns of CP observed in many cases.
\input{fig01.tex}

\input{fig02.tex}

\subsection{Luminosities versus Circular Polarization} 

Table 2 shows a summary of the CP detections for this and some previous studies.  
The table includes high-, intermediate-, and low-mass star-forming regions 
and orders the sources in terms of decreasing luminosities.
\input{tab1.tex}

\input{tab2.tex}

The most striking correlation we have found is that 
the observed maximum CP increases with increasing luminosity of the illuminating YSOs (Figure 3). 
The luminosities of YSOs are a measure of their stellar masses. 
In fact, the mass estimate for these kinds of YSOs 
(Class I sources and intermediate-mass or massive protostars) is difficult 
and we use the luminosities for our discussion.
This correlation could support the radiative torque mechanism for the alignment of dust grains (e.g., Lazarian \& Hoang 2007; Matsumura \& Bastien 2009). 
The torque amplitude is proportional to the radiation intensity 
and the most efficient alignment is for grains larger than $\lambda/2$ 
(as might be expected to occur in star-forming regions). 
\input{fig03.tex}

Alternatively, this apparent correlation can be simply understood as being due to a stronger magnetic field in dense massive star-forming regions producing better dust alignment and thus more readily producing CPs.
To check if the magnetic fields in the observed regions are in fact strong, we have investigated Zeeman measurements toward our observed regions and found that two high-mass star-forming regions, OMC-1 and NGC 6334-V, show strong magnetic fields. Crutcher et al. (1999, 2010) reported a magnetic field strength of 360 $\mu$G in the OMC-1 region from the CN Zeeman effect. Sarma et al. (2000) reported a magnetic field strength of 150 $\mu$G in the NGC 6334 complex to the east of the NGC 6334-V region from the OH Zeeman effect. Unfortunately, there is no further information on the magnetic field strength directly observed in the intermediate- and low-mass star-forming regions OMC-2, OMC-3, Cha IRN, and L1551. However, the survey of Crutcher et al. (2010) implies that the magnetic field strengths are typically weak ($< 30$ $\mu$G) in the intermediate- and low-mass star-forming regions ($< n_{\rm H} \sim 1 \times 10^3$ cm$^{-3}$). 

In either case 
the positive correlation in Figure 3 is consistent with dichroic extinction
being responsible for the higher values of CP
as the higher mass star-forming regions are associated with higher extinction.

\subsection{Color versus Circular Polarization}

We examined empirical relationship between the $K_s$-band CP and the $H - K_s$ color for the points along the two diagonals passing through the centers of the quadrupolar CP patterns for OMC-1 and NGC 6334-V (the figure not shown here). 
We found that the highest degrees of CP tend to be associated with the reddest colors. 
In addition, the slope of the CP change with color is also dependent on the clouds 
(slopes are $\sim$13.7 and $\sim$6.0 \%/mag for OMC-1 and NGC 6334-V, respectively). 
Unsurprisingly, then, that the environments for dust grain alignments 
such as radiation field or magnetic field strength 
are different in each star forming region. In previous studies, 
Fukue et al. (2009) showed that 
there is a good correlation between CP (i.e., Stokes $V$) and $H - K_s$ color, 
with the latter used as a proxy for extinction. In addition, 
Kwon et al. (2013) showed that high CP may be produced by scattering from the IRN 
followed by dichroic extinction by an optically thick foreground cloud containing aligned dust grains. 
These results again are consistent with dichroic extinction being the major contributor 
to the high degrees of CP found in massive star-forming regions.

\subsection{Large and Extended Circular Polarization} 

The extent of the infrared CP regions is large ($\sim$0.1 to $\sim$1 pc), 
which has potentially important implications for the surrounding regions and any neighboring YSOs.  
Since many young stars are born with a high spatial density of $\sim$10$^2$--10$^3$ stars pc$^{-2}$  
(e.g., Lada et al. 2004), 
such extensive and high degrees of CP from massive and young stellar objects can easily affect molecules in circumstellar structures around nearby low-mass YSOs if the CP is also relatively high not only at IR but also at UV wavelengths.  

This provides a possible explanation for the origin of the homochirality of amino acids and sugars 
in all terrestrial organisms.  Bailey et al. (1998) (and see Fukue et al. 2009 and Kwon et al. 2013), 
proposed that an enantiomeric excess could be produced by asymmetric photolysis of any chiral prebiotic molecules by the action of circularly polarized light produced in high-mass star-forming regions.

\section{CONCLUSIONS}

We have presented the first systematic survey of a combination of linear and circular polarimetry 
in six star-forming regions, covering high-mass, intermediate-mass, and low-mass YSOs;  
OMC-1, NGC 6334-V, OMC-2, OMC-3, Cha IRN, and L1551 regions. 
We have investigated the polarization properties of each region with their morphology, such as patterns, extents, and maximum degrees of LP and CP. Large ($> 2$ \%) infrared CP is universally observed in IRNe.  
In particular, extremely large ($> 10$ \%) CP values are observed in high-mass star-forming regions, 
and with extents up to $\sim 1$ pc. Such a large CP is most readily explained by dichroic extinction 
and the correlation with the luminosity of the sources supports the radiative torque mechanism 
for aligning the dust grains.  
Such extended regions, together with the high degrees of CP, 
have important implications for the environments of massive YSOs, 
including any prebiotic chiral molecules.

\acknowledgments

J.K. was supported by Grants-in-Aid for JSPS Fellows ($24 \times 110$ and $26 \times 04023$).
M.T. was supported by MEXT KAKENHI grant numbers 19204018 and 22000005. 
The IRSF project is a collaboration between Nagoya University and the
South African Astronomical Observatory (SAAO) supported by the
Grants-in-Aid for Scientific Research on Priority Areas (A) (no.
10147207 and no. 10147214) and Optical \& Near-Infrared Astronomy
Inter-University Cooperation Program, from the Ministry of
Education, Culture, Sports, Science and Technology (MEXT) of Japan and
the National Research Foundation (NRF) of South Africa.

\end{document}

%% file: fig01.tex
\begin{figure*}
\epsscale{2.0}
\plotone{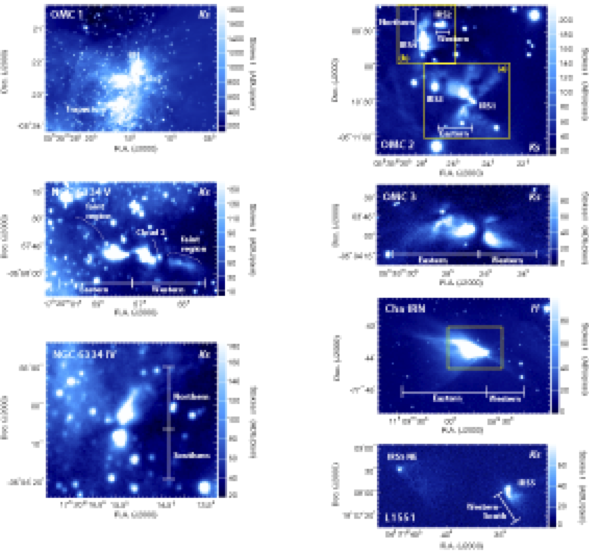}
\vspace{-0.5\baselineskip}
\centerline{\scriptsize }
\vspace{-0.5\baselineskip}
\caption{
Finding charts of the surveyed regions on the Stokes $I$ images observed from the
IRSF 1.4 m telescope. The yellow boxes correspond to the CP regions
shown in Figure 2 for OMC 2 and Cha IRN.
The color scaling is linear and in ADU pixel$^{-1}$.
}
\end{figure*}

%% file: fig02.tex
\begin{figure*}
\epsscale{2.0}
\plotone{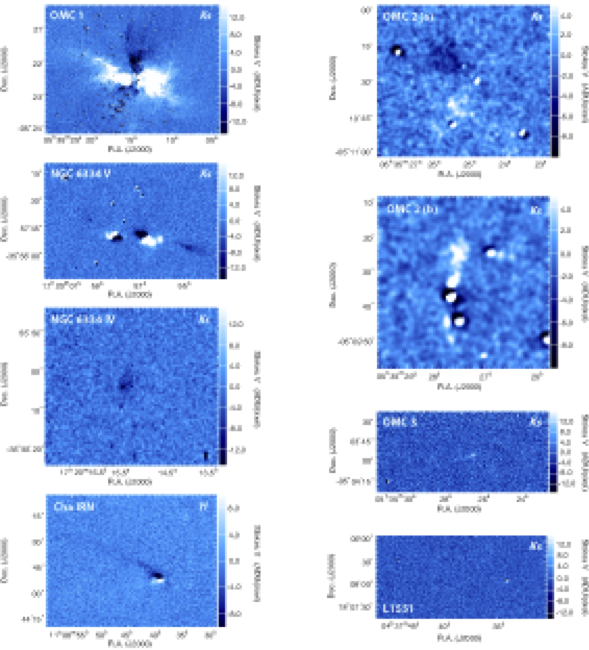}
\vspace{-0.5\baselineskip}
\centerline{\scriptsize }
\vspace{-0.5\baselineskip}
\caption{
Stokes $V$ images observed from the IRSF 1.4 m telescope. OMC 2(a), OMC 2(b), and Cha IRN show marked regions with yellow boxes in Figure 1.
The white and black colors indicate positive and negative Stokes $V$ (in ADU pixel$^{-1}$), respectively, and the blue background indicates zero levels.
}
\end{figure*}

%% file: tab1.tex
\begin{table*}[hbtp]
\begin{center}
\footnotesize
\begin{threeparttable}
\caption{Properties of Circular and Linear Polarization}
  \begin{tabular}{lllr}
\hline
\hline
Object						&	Pattern 				&   Extent 				&	$|$ Maximum Degree $|$ \\ \hline
\multicolumn{4}{c}{ Properties of Circular Polarization } \\ \hline
OMC-1 IRN (IRc2)				&	Asymmetric quadropolar	&   0.45 pc (210\arcsec)		&	16.2 $\pm$ 0.2 \% 				\\
NGC 6334-V IRN 				&	Asymmetric quadropolar &   0.65 pc (80\arcsec)		&	19.8 $\pm$ 0.5 \% 				\\	
NGC 6334-IV IRN 				&	Negative bipolar		&   0.33 pc (40\arcsec)		& 	3.5 $\pm$ 0.1 \% 				\\	
OMC-2 IRN (IRS 1)				&	Quadropolar			&   0.17 pc (80\arcsec)		& 	4.6 $\pm$ 0.4 \% 		\\	
OMC-2 IRN (IRS 2)				&	Positive monopolar		&   0.04 pc (20\arcsec)		& 	1.8 $\pm$ 0.2 \% 		\\	
OMC-2 IRN (IRS 4)				&	Positive monopolar		&   0.09 pc (40\arcsec)		& 	4.3 $\pm$ 0.2 \% 		\\	
OMC-3 IRN (IRAS 05329-0505)	&	Positive bipolar		&   0.22 pc (100\arcsec)		& 	1.9 $\pm$ 0.2 \% 		\\
Cha IRN (IRAS 11072-7727)		&	Asymmetric quadropolar	&   0.12 pc (150\arcsec)		& 	1.1 $\pm$ 0.1 \% 		\\
L1551 IRN (IRS 5)				&	Positive monopolar		&   0 02 pc (30\arcsec)		& 	1.6 $\pm$ 0.2 \% 		\\	
L1551 IRN (IRS 5 NE)			&	Negative monopolar		&   0.007 pc (10\arcsec)		& 	1.1 $\pm$ 1.0 \% 		\\	
\hline
\multicolumn{4}{c}{ Properties of Linear Polarization } \\ \hline
OMC-1 IRN (IRc2)				&	Centrosymmetric  		&   0.91 pc (420\arcsec)	&	 30 \%  				\\
NGC 6334-V IRN 				&	Asymmetric bipolar 		&   0.65 pc (80\arcsec)	&	 70 \%  				\\	
NGC 6334-IV IRN 				&	Bipolar 				&   0.49 pc (60\arcsec)	&	 50 \%  				\\	
OMC-2 IRN (IRS 1)				&	Centrosymmetric		&   0.22 pc (100\arcsec)	&	 30 \%  				\\	
OMC-2 IRN (IRS 2)				&	Bipolar 				&   0.05 pc (25\arcsec)	&	 20 \%  				\\	
OMC-2 IRN (IRS 4)				&	Bipolar 				&   0.13 pc (60\arcsec)	&	 30 \%  				\\	
OMC-3 IRN (IRAS 05329-0505)	&	Asymmetric bipolar		&   0.43 pc (200\arcsec)	&	 70 \%  				\\
Cha IRN (IRAS 11072-7727)		&	Bipolar				&   0.32 pc (420\arcsec)	&	 35 \%  				\\
L1551 IRN (IRS 5)				&	Monopolar			&   0.17 pc (240\arcsec)	&	 30 \%  				\\	
L1551 IRN (IRS 5 NE)			&	Monopolar			&   0.09 pc (120\arcsec)	&	 15 \%  				\\	
\hline
\multicolumn{4}{l}{Note. -- CPs were estimated in the $K_s$ band except for Cha IRN estimated in the $H$ band. }\\
\multicolumn{4}{l}{Distances are referred to in (1)--(4).} \\
\multicolumn{4}{l}{(1) OMC is at a distance of 450 pc (see, e.g., Genzel \& Stutzki 1989). (2) NGC 6334-V is at a } \\
\multicolumn{4}{l}{ distance of 1700 pc (Neckel 1978).  (3) Cha IRN is at a distance of 160 pc (Whittet et al. 1997).  }\\
\multicolumn{4}{l}{ (4) L1551 is at a distance of 150 pc (Kenyon et al. 1994; Reipurth 1999). }\\
\multicolumn{4}{l}{ Note that extents are approximate values. For some regions, each lobe is separately discussed.}\\
\label{table:properties}
\end{tabular}
\end{threeparttable}
\end{center}
\end{table*}

%% file: tab2.tex
\begin{table*}[hbtp]
\begin{center}
\footnotesize
\begin{threeparttable}
\caption{Summary of CP Detections of This and Previous Studies (Sorted by Luminosity)} 
\begin{tabular}{llrrlc}
\hline
\hline
Object (Illuminating Source)		&  Mass				&	Luminosity 					&  $|$CP$|$				& Extent 	&  Note.	\\ 
							&  					&	 ($L_{\sun}$)					&  (\%)				&  (pc)	&  		\\ \hline
NGC 6334-V IRN 			 	&  High				&	$\sim$2 $\times$ 10$^5$ \,	(1)	&  20 	 			& 0.65 	&  $^a$    \\
OMC-1 IRN (IRc2)				&  High				&	$\sim$1 $\times$ 10$^5$ \,	(2)	&  16 	 			& 0.45	&  $^a$   	\\
HH 135--136					&  High/intermediate	&	$\sim$1 $\times$ 10$^4$ \,	(3)	&  15 				& 0.52	&  $^b$	\\
OMC-2 IRN (IRS 4)				&  Intermediate		&	$\sim$3 $\times$ 10$^2$ \,	(4)	&  4 		  			& 0.09	&  $^a$    \\
R CrA						&  Intermediate		&	$\sim$1 $\times$ 10$^2$ \,	(5)	&  5 					& 0.012	&  $^c$	\\
OMC-2 IRN (IRS 2)				&  Intermediate		&	$\sim$60				   \,	(4)	&  2 		  			& 0.04 	&  $^a$    \\
OMC-3 IRN (IRAS 05329-0505) 	&  Intermediate		&	$\sim$50			 	   \,	(5)	&  2 		 			& 0.22	&  $^a$    \\
OMC-2 IRN (IRS 1)				&  Intermediate		&	$\sim$40			 	   \,	(4)	&  5 		  			& 0.17 	&  $^a$    \\
L1551 IRN (IRS 5)				&  Low				&	$\sim$15.3			   \,	(6)	&  1.6	 		 	& 0.02	&  $^a$    \\
Cha IRN (IRAS 11072-7727)		&  Low				&	$\sim$14.4			   \,	(7)	&  1.1				& 0.12	&  $^a$	\\ 
L1551 IRN (IRS 5 NE)			&  Low				&	$\sim$4.2			   \,	(6)	&  $<$ 2 				& 0.007	&  $^a$    \\
GSS 30						&  Low				&	$\sim$1				   \,	(8)	& 1.7 				& 0.02	&  $^d$	\\
$^\dagger$NGC 6334-IV IRN		&  \dots				&	\dots						& 3		 			& 0.65 	&  $^a$    \\\hline
\multicolumn{6}{l}{Note. --  CPs were estimated in the $K$ or $K_s$ band except for R CrA and Cha IRN estimated in the $H$ band.} \\
\multicolumn{6}{l}{Luminosities are referred to in (1)--(8). } \\
\multicolumn{6}{l}{(1)  Harvey \& Gatley 1983; Loughran et al. 1986; McBreen et al. 1979  \, (2)  Gezari et al. 1998 } \\
\multicolumn{6}{l}{(3)  Gredel 2006; Ogura \& Walsh 1992  \, (4) Adams et al. 2012 \, (5) Wilking et al. 1986} \\
\multicolumn{6}{l}{(6) Ainsworth et al. 2012 \, (7) Ageorges et al. 1996 \, (8) Zhang et al. 1997} \\
\multicolumn{6}{l}{$^\dagger$The luminosity of NGC 6334-IV IRN is unknown.} \\
\multicolumn{6}{l}{$^a$ This work \, $^b$ Chrysostomou et al. (2007) \, $^c$  Clark et al. (2000)  \, $^d$ Chrysostomou et al. (1997)} \\
\label{table:CPsummary}
\end{tabular}
\end{threeparttable}
\end{center}
\end{table*}

%% file: fig03.tex
\begin{figure*}
\epsscale{1.5}
\plotone{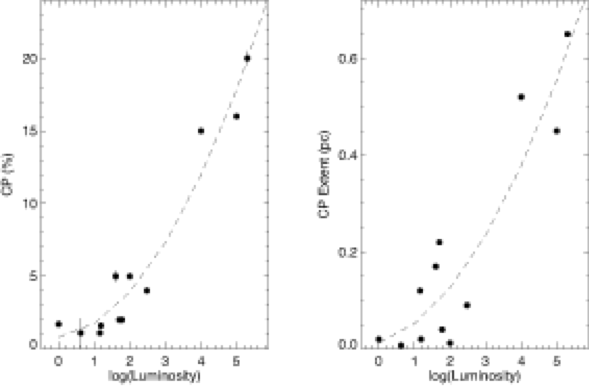}
\vspace{-0.5\baselineskip}
\centerline{\scriptsize }
\vspace{-0.5\baselineskip}
\caption{
Relationship between luminosities and CP.
Left panel: luminosity ($L_{\odot}$) vs. maximum degree of CP. 
Right panel: luminosity ($L_{\odot}$) vs. CP extent (pc).
Dashed line: curve fitting by second-order polynomial function. Note however that this is just an empirical fitting.
}
\end{figure*}